\documentclass{article}

\usepackage{arxiv}

\usepackage[utf8]{inputenc} 
\usepackage[T1]{fontenc}    
\usepackage{url}            
\usepackage{booktabs}       
\usepackage{amsfonts}       
\usepackage{amsmath}
\usepackage{nicefrac}       
\usepackage{microtype}      
\usepackage{graphicx}
\usepackage{doi}
\usepackage{caption,subcaption}
\usepackage{enumerate}
\usepackage{xcolor}
\usepackage{xfrac}
\usepackage{afterpage}

\newcommand{\acknowledgement}
              [1]
              {
               \bgroup
               \flushleft
               \small\bf
               #1
               \par
               \egroup
              }

\title{Modeling of 2D self-drifting flame-balls in Hele-Shaw cells}

\author{ {Jorge Yanez}\thanks{Corresponding author} \\
	Institute for Thermal Energy Technology and Safety\\
        Karlsruhe Institute of Technology\\
        Hermann-von-Helmholtz-Platz, 1\\
        Karlsruhe, Germany.\\
	\texttt{jorge.yanez@kit.edu} \\
	\And
	Leonid Kagan\\
	School of Mathematical Sciences,\\
        Tel Aviv University\\
        Tel Aviv, Israel.\\
	\texttt{kaganleo@tauex.tau.ac.il} \\
	\AND
        Mike Kuznetsov\\
        Institute for Thermal Energy Technology and Safety\\
        Karlsruhe Institute of Technology\\
        Hermann-von-Helmholtz-Platz, 1\\
        Karlsruhe, Germany.\\
	\texttt{mike.kuznetsov@kit.edu}
	\And
	Gregory Sivashinsky \\
        School of Mathematical Sciences,\\
        Tel Aviv University\\
        Tel Aviv, Israel.\\
	\texttt{grishas@tauex.tau.ac.il} \\
}
\date{}

\renewcommand{\shorttitle}{On disintegration of lean hydrogen flames in narrow gaps}

\begin{document}

\title{\LARGE On disintegration of lean hydrogen flames in narrow gaps
}

\maketitle

\begin{abstract} 
   The disintegration of near limit flames propagating through the gap of Hele-Shaw cells has recently become a subject of active research.  In this paper, the flamelets resulting from the disintegration of the continuous front are interpreted in terms of the Zeldovich flame-balls stabilized by volumetric heat losses.  A complicated free-boundary problem for 2D self-drifting near circular flamelets is reduced to a 1D model.  The 1D formulation is then utilized to obtain the locus of the flamelet velocity, size, heat losses and Lewis numbers at which the self-drifting flamelets may exist.

\end{abstract}



\clearpage

\section{Introduction}

\begin{figure}[h!]

     \begin{subfigure}[b]{0.32\columnwidth}
         \centering
         \vspace{10pt}
         \includegraphics[width=\columnwidth]{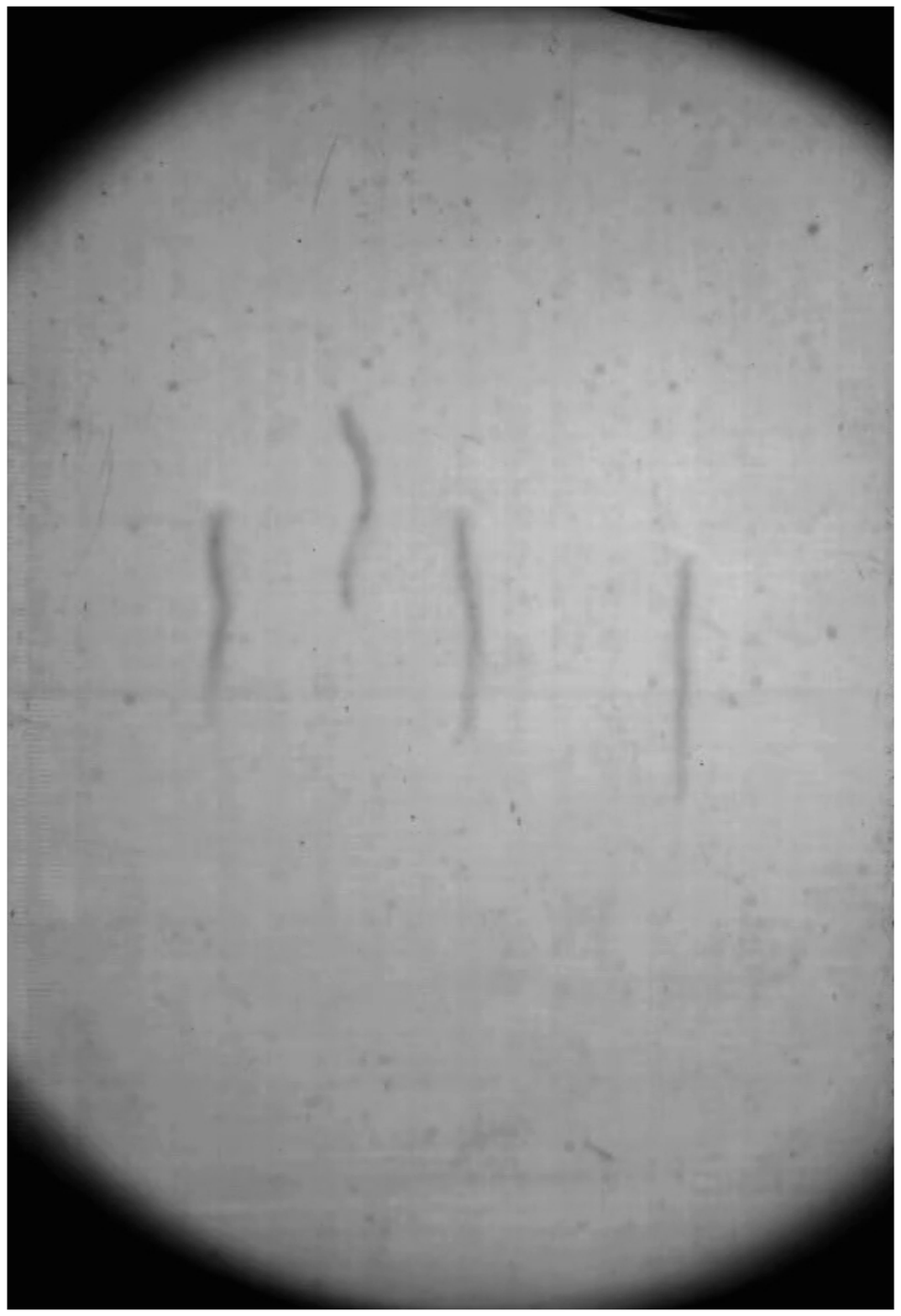} 
         \caption{}
         \label{fig:1f}
     \end{subfigure}
     \begin{subfigure}[b]{0.32\columnwidth}
         \centering
         \vspace{10pt}
         \includegraphics[width=\columnwidth]{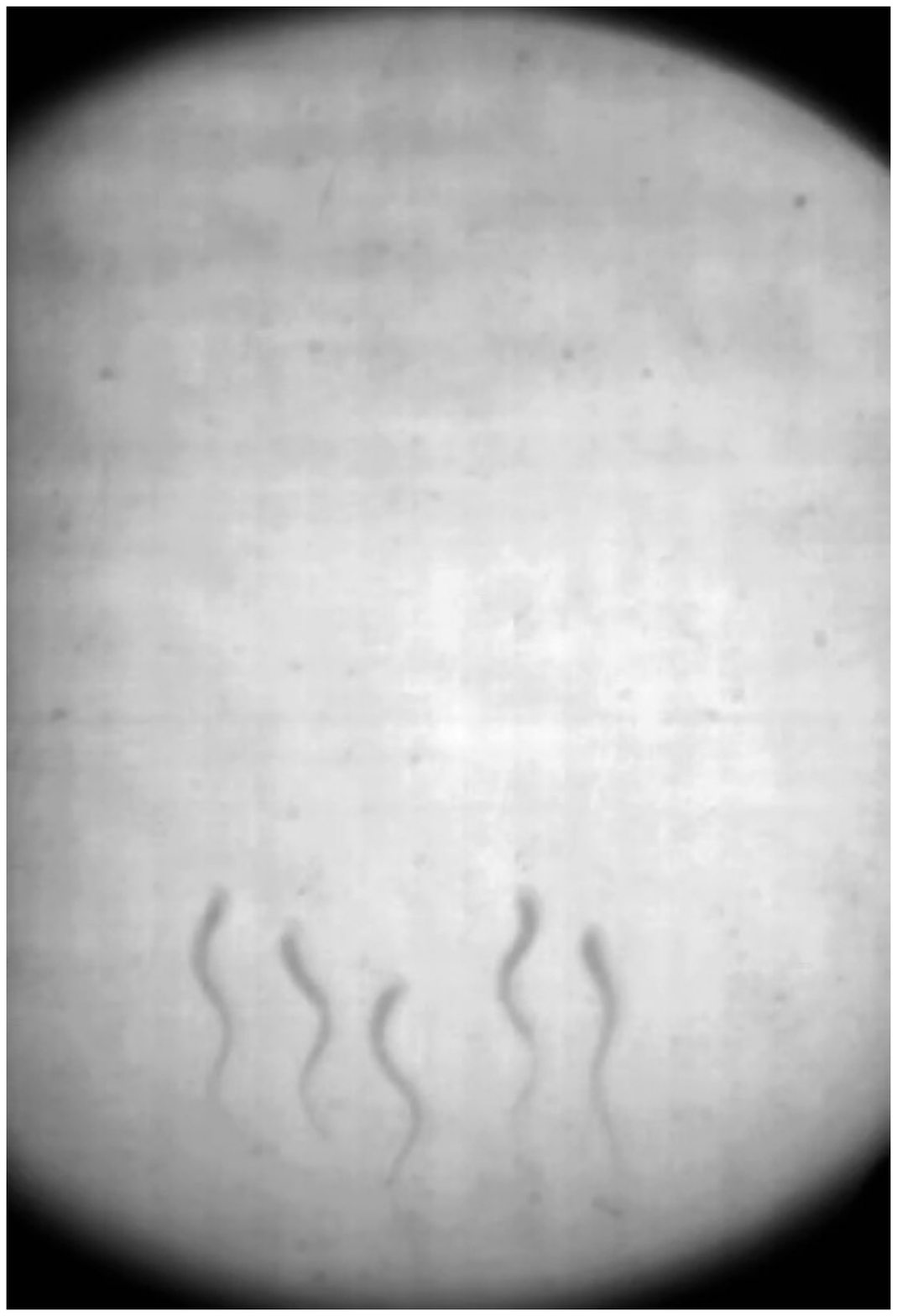} 
         \caption{}
         \label{fig:1e}
     \end{subfigure}
     \begin{subfigure}[b]{0.32\columnwidth}
         \centering
         \includegraphics[width=\columnwidth]{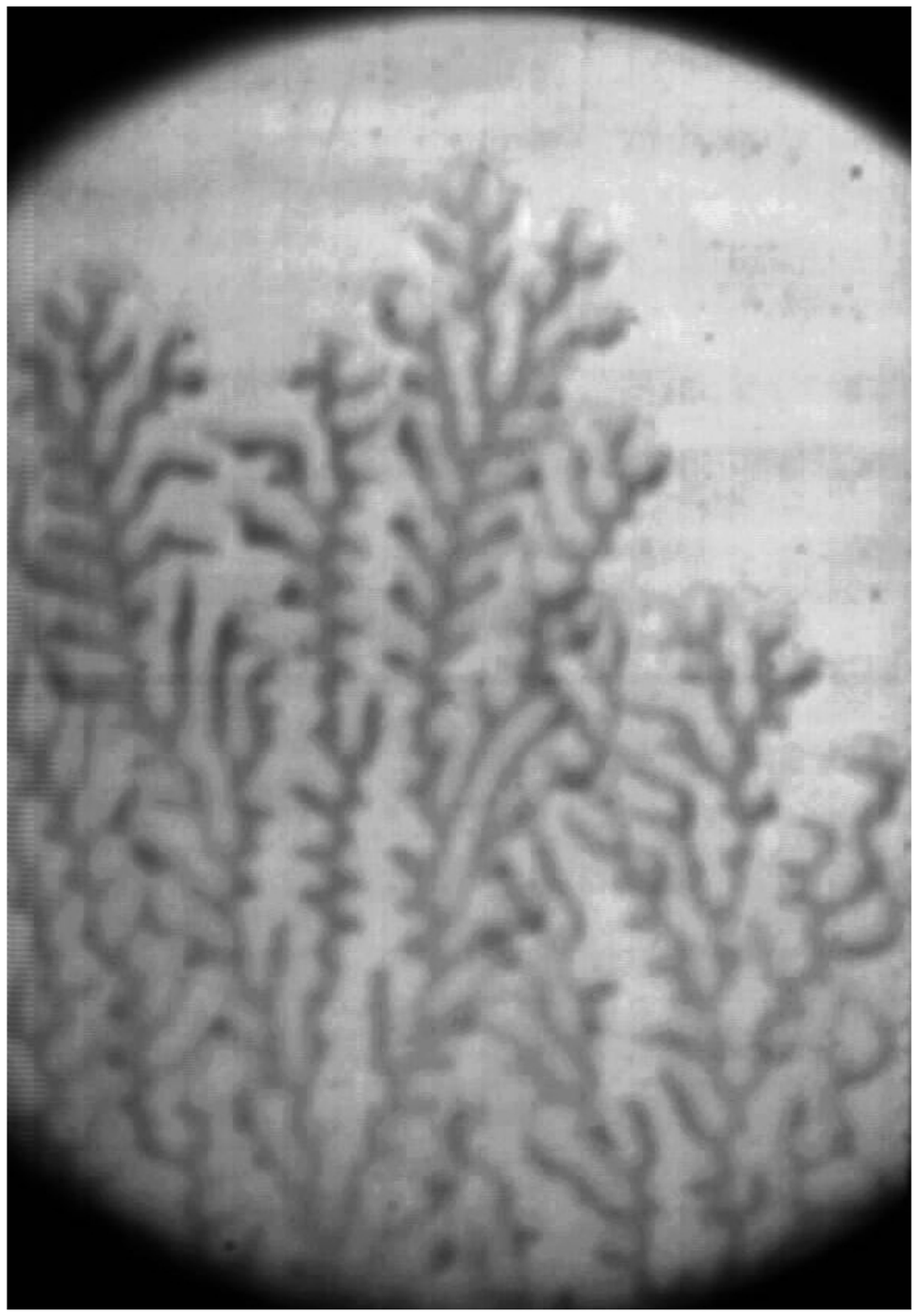}
         \caption{}
         \label{fig:1a}         
     \end{subfigure}

     \centering
     \begin{subfigure}[b]{0.32\columnwidth}
         \centering
         \vspace{10pt}
         \includegraphics[width=\columnwidth] {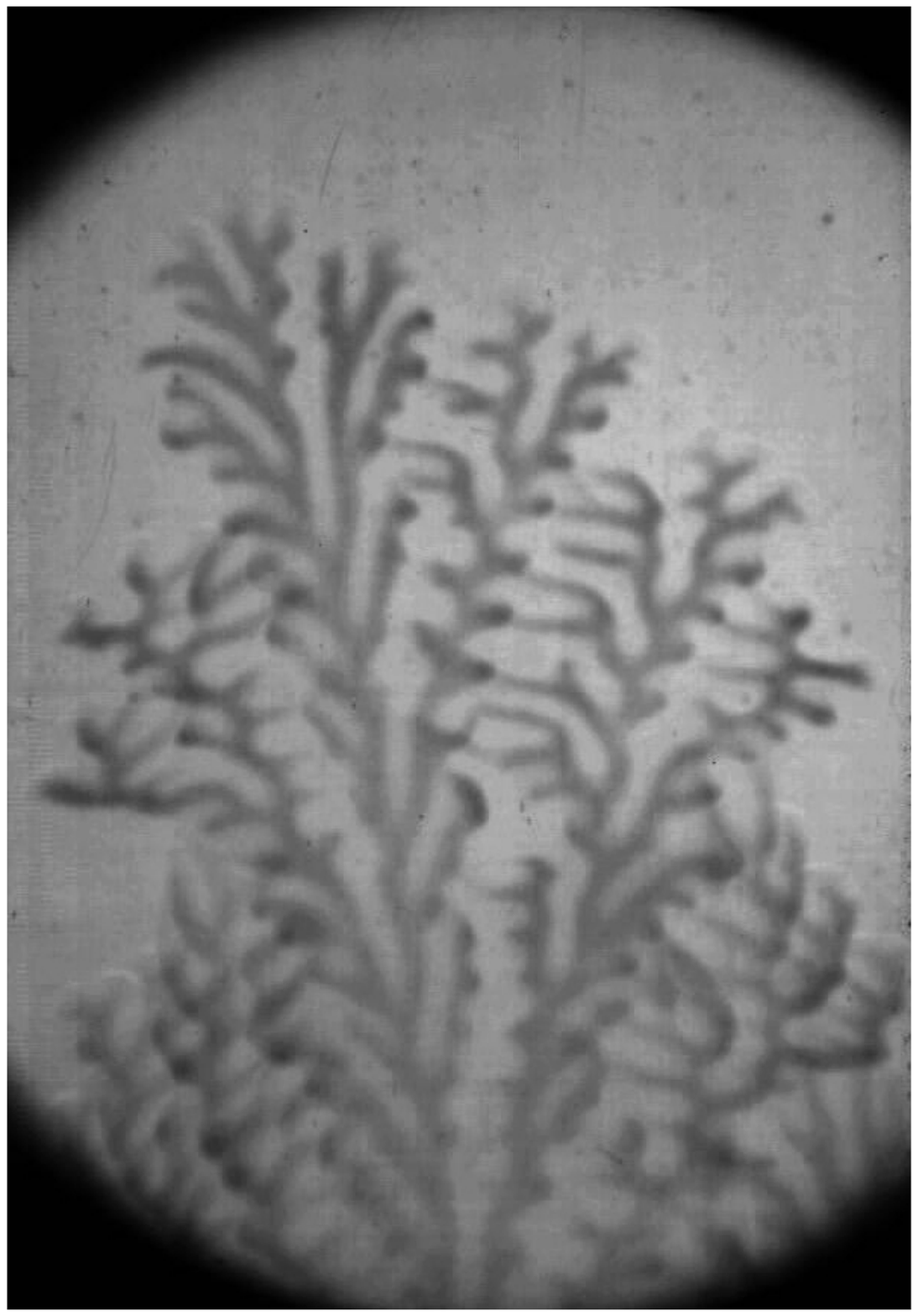}      
         \caption{}
         \label{fig:1b}
     \end{subfigure}%
     \begin{subfigure}[b]{0.32\columnwidth}
         \centering
         \includegraphics[width=\columnwidth]{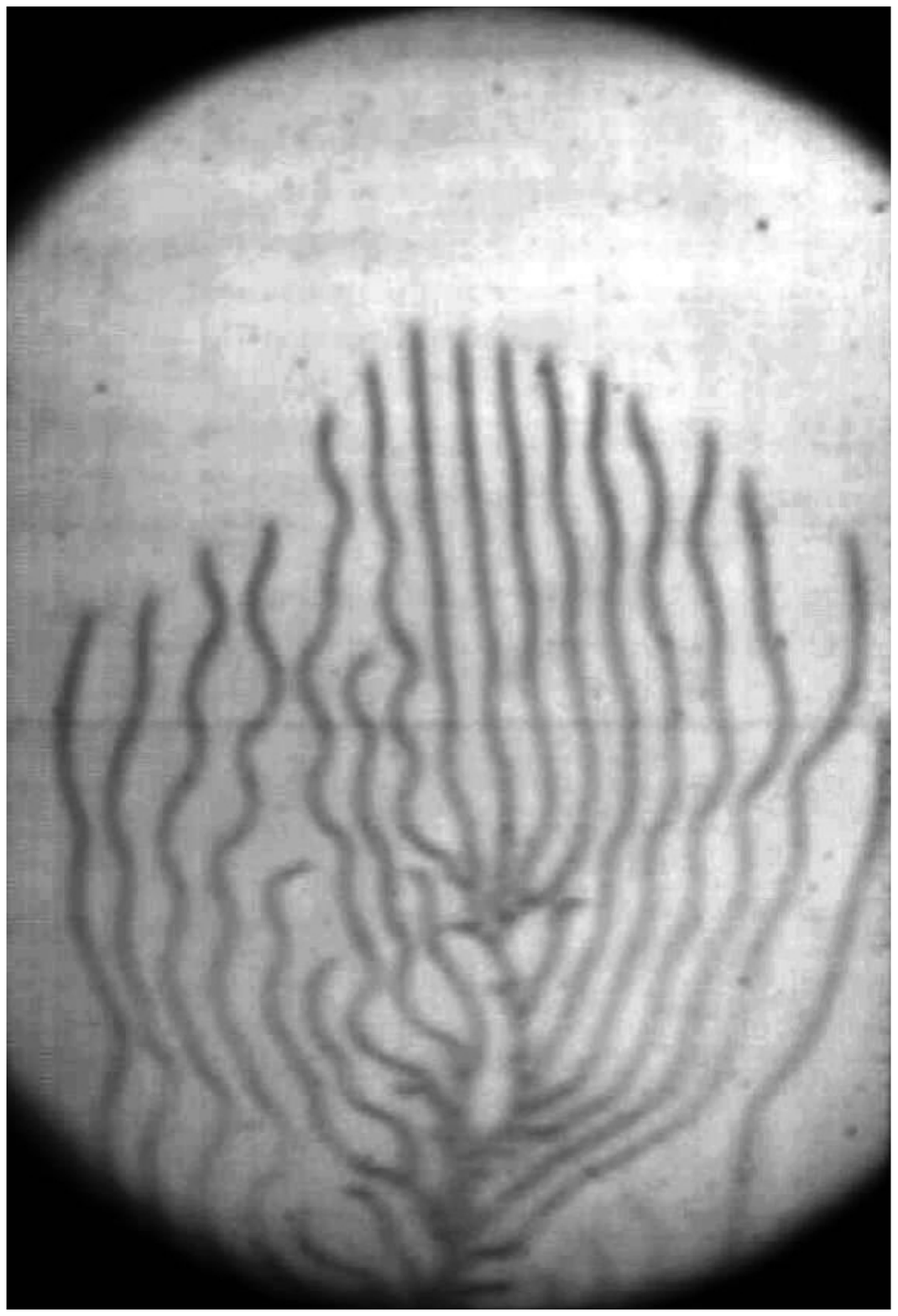}
         \caption{}
         \label{fig:1c}         
     \end{subfigure}%
     \begin{subfigure}[b]{0.32\columnwidth}
         \centering
         \vspace{10pt}
         \includegraphics[width=\columnwidth]{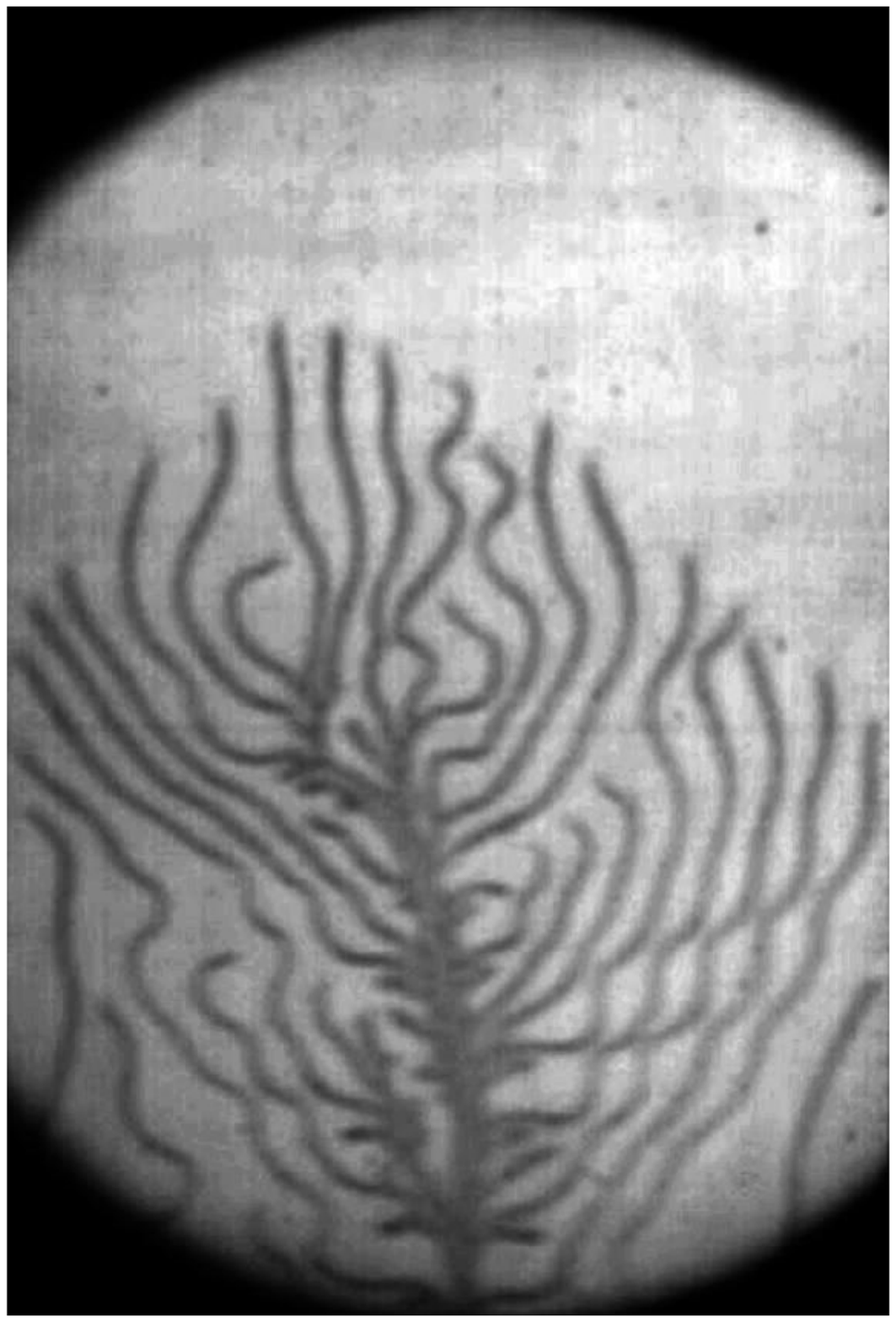}    
         \caption{}
         \label{fig:1d}
     \end{subfigure} 

     \vspace{1pt}
 \caption{Hydrogen-air flamelets propagating through narrow gaps: 
 (a) $d=5mm$, $\% H_2 = 5.0$; 
 (b) $d=3mm$, $\% H_2 = 7.5$; 
 (c) $d=4mm$, $\% H_2 = 7.0$; 
 (d) $d=5mm$, $\% H_2 = 6.0$; 
 (e) $d=2mm$, $\% H_2 = 9.75$;
 (f) $d=2mm$, $\% H_2 = 9.75$.
   Regimes: (a)(b) one-headed finger, (c)(d)(e)(f)  one-headed branching.}
	\label{kk1}
\end{figure}

Hydrogen flames have gained increasing attention over the last years as a clean and efficient energy solution [1].
Apart from their technological relevance, hydrogen flames are remarkably rich dynamically.  Because of the high diffusion coefficient of molecular hydrogen, lean hydrogen-air flames are known to experience diffusive-thermal
instability [2,3] manifesting themselves in the formation of a cellular structure in a state of chaotic self-motion.  Moreover, as has recently been discovered by Veiga-López et al. [4], lean hydrogen-air flames evolving in narrow gaps of Hele-Shaw chambers, and where heat-loss effects become important, continuous cellular flames may disintegrate forming self-drifting cup-like flamelets leaving tree-like or sprout-like traces (Fig. \ref{kk1}). 

\clearpage

   These unexpected propagation modes extend flammability limits beyond those of the planar flames. This is particularly important for the safety studies of hydrogen-powered devices [5] as it implies that hydrogen flames may propagate in gaps much narrower than initially anticipated.
   
\section{Reaction-diffusion model}\addvspace{10pt}
\label{sec:two}
Some of the above features of the Hele-Shaw flames may be successfully captured by an ultra-simple constant-density, buoyancy-free,
reaction-diffusion model [6]. In suitably scaled variables and parameters it reads,
\begin{align}
	&\dfrac{\partial T}{\partial t} 
	=
	\dfrac{\partial^2 T }{\partial x^2}
	+
	\dfrac{\partial^2 T }{\partial y^2}
	+(1-\sigma)\Omega
	-
	q(T-\sigma),\label{eq:eq1}
	\\
	&\dfrac{\partial C}{\partial t} 
	=
	\dfrac{1}{Le}
	\left(
	\dfrac{\partial^2 C }{\partial x^2}
	+
	\dfrac{\partial^2 C }{\partial y^2}
	\right) 
	-
	\Omega,\label{eq:eq2}
	\\
	&\Omega 
	=
 \dfrac{1}{2Le}(1-\sigma)^2N^2C\exp{\left[N\left(1-\dfrac{1}{T}\right)\right]},
\end{align}
where	$T=$ temperature in units of $T_b$, the adiabatic temperature of combustion products;    $C=$ concentration of the deficient reactant in units of $C_0$, its value in the fresh mixture; $x,y,t=$ spatio-temporal coordinates in units of 	$\ell_{th}=D_{th}/U_b$ and $\ell_{th}/U_b$, respectively;	$\ell_{th}=$ thermal width of the flame;	$U_b=$ speed of a planar adiabatic flame;	  $D_{th} =$ thermal diffusivity of the mixture,	$\sigma=T_0/T_{b}$, where	$T_0=$ temperature sustained at the walls of the Hele-Shaw cell; 
$Le=D_{th}/D_{mol}$ is the Lewis number;
$D_{mol}=$ molecular diffusivity of the deficient reactant;		$N=T_a/T_b=$ scaled activation energy;    $T_a   =$ activation temperature;	 $\Omega =$ appropriately normalized reaction rate to ensure that at large $N$       the scaled speed of the planar adiabatic flame is close to unity; $q=$ scaled heat loss intensity specified as	$(\pi\ell_{th}/d)^2$;
$d=$ width of the Hele-shaw cell. The adopted expression for $q$ stems from the 1D heat equation, $T_t=T_{zz}$, considered over the Hele-Shaw gap and subjected to isothermal boundary conditions, $T(z=\pm d/2 \ell_{th}, t)=\sigma$. Hence, $T-\sigma\sim\exp{\left[-\left(\pi \ell_{th}/d\right)^2 t\right]}\cos{\left[\left(\pi \ell_{th}/d\right)z\right]}$. The exponential rate of the temperature decay is then extrapolated over the quasi-2D formulation of \eqref{eq:eq1}-\eqref{eq:eq2} in the $(x,y)$ plane.

\begin{figure}[h!]
	\centering
     \begin{subfigure}[b]{0.32\columnwidth}
         \centering
         \includegraphics[width=\columnwidth]{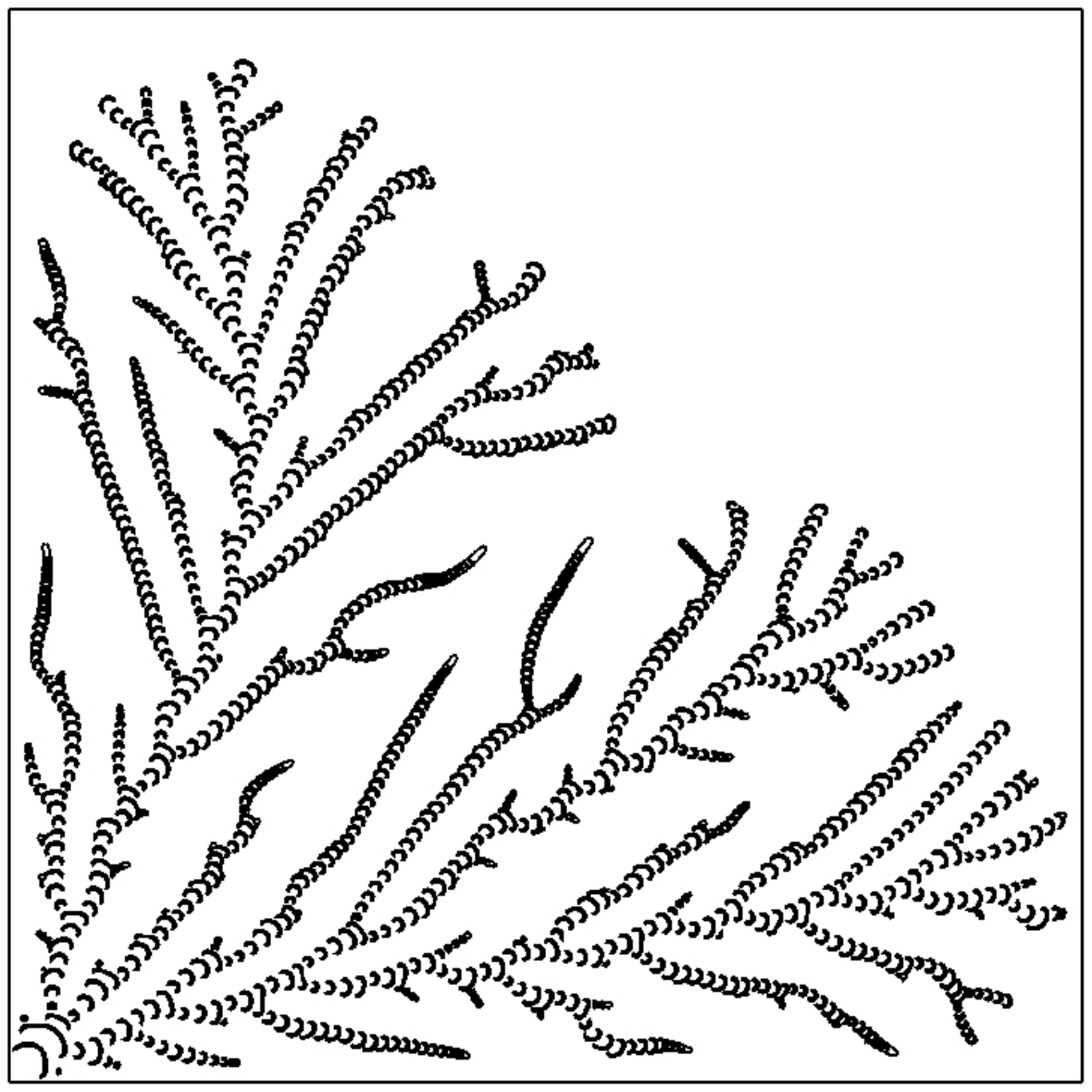}
         \caption{}
         \label{fig:2a}         
     \end{subfigure}
     \begin{subfigure}[b]{0.32\columnwidth}
         \centering
         \vspace{10pt}
         \includegraphics[width=\columnwidth]{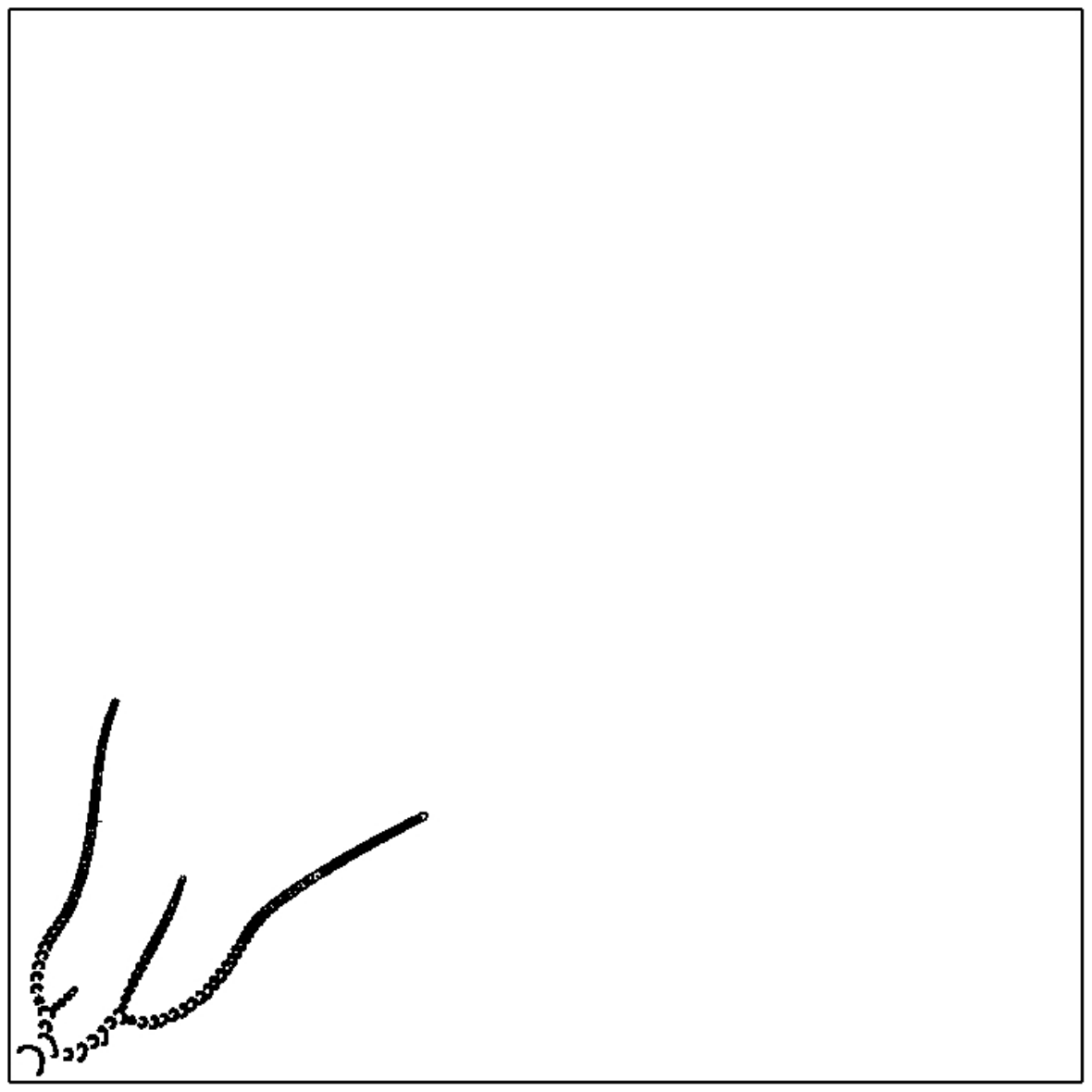}
        
         \caption{}
         \label{fig:2b}
     \end{subfigure}  
     \vspace{1pt}
 \caption{Reaction rate $\Omega=0.4$ at consecutive equispaced instants for one-headed flamelets under heat losses.  Simulation of the reactive-diffusive system (1)-(3) for $N=10$; $Le=0.25$; $\sigma=0.2$; $q=0.14$, $0<t<740$ (a); $q=0.20$, $0<t<1000$ (b); $0<x,y<480$. Initial conditions, $T(x,y,0)=\sigma+(1-\sigma)\exp{[-(x^2+y^2)/\ell^2]}
 $, $\ell=0.25$, $C(x,y,0)=1$,  and adiabatic boundary conditions.}
	\label{kk2}
\end{figure}

\begin{figure}[h!]
	\centering
     \begin{subfigure}[b]{0.32\columnwidth}
         \centering
         \includegraphics[width=\columnwidth]{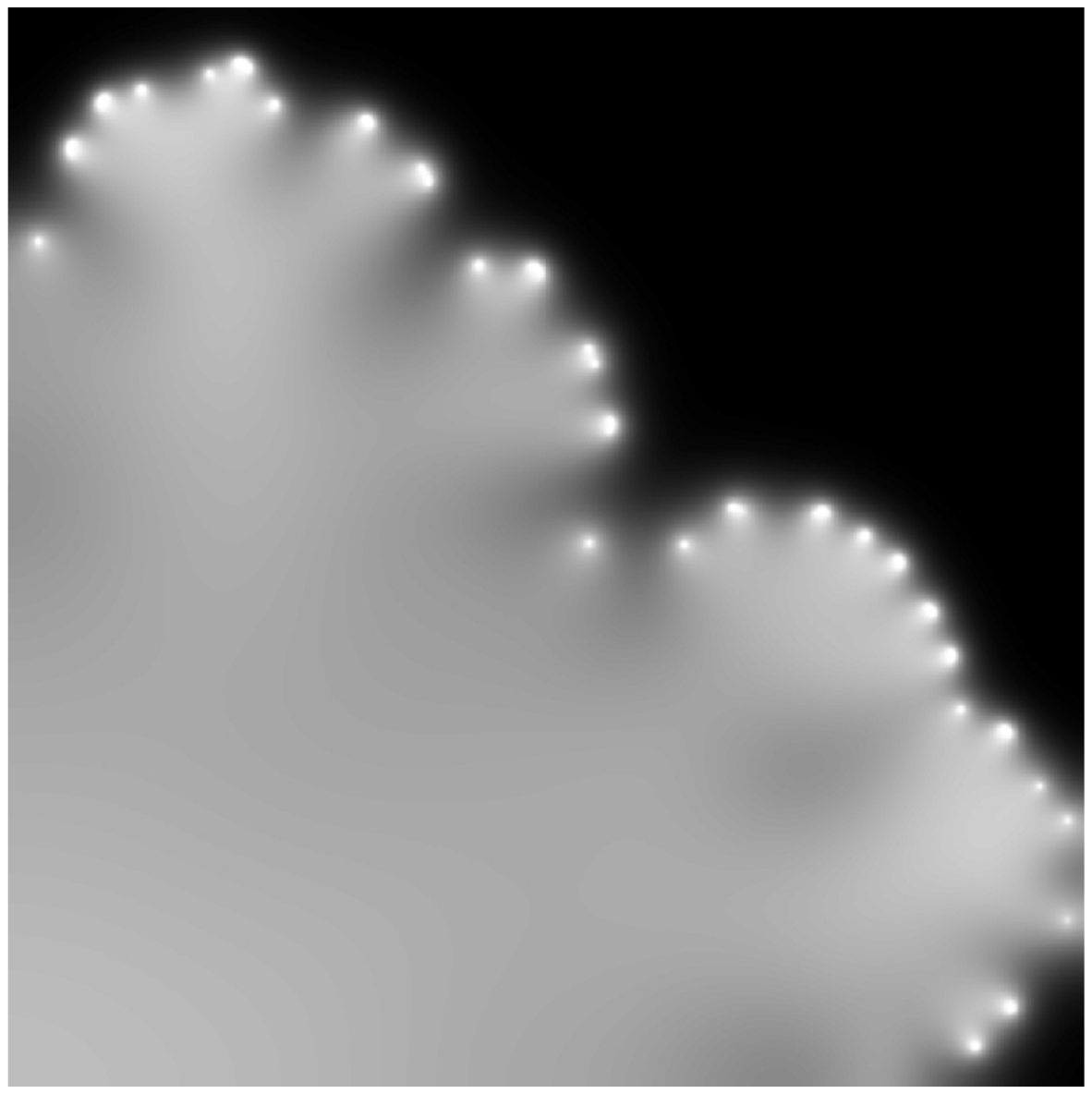}
         \caption{}
         \label{fig:3a}
     \end{subfigure}
     \begin{subfigure}[b]{0.32\columnwidth}
         \centering
         \vspace{10pt}
         \includegraphics[width=\columnwidth]{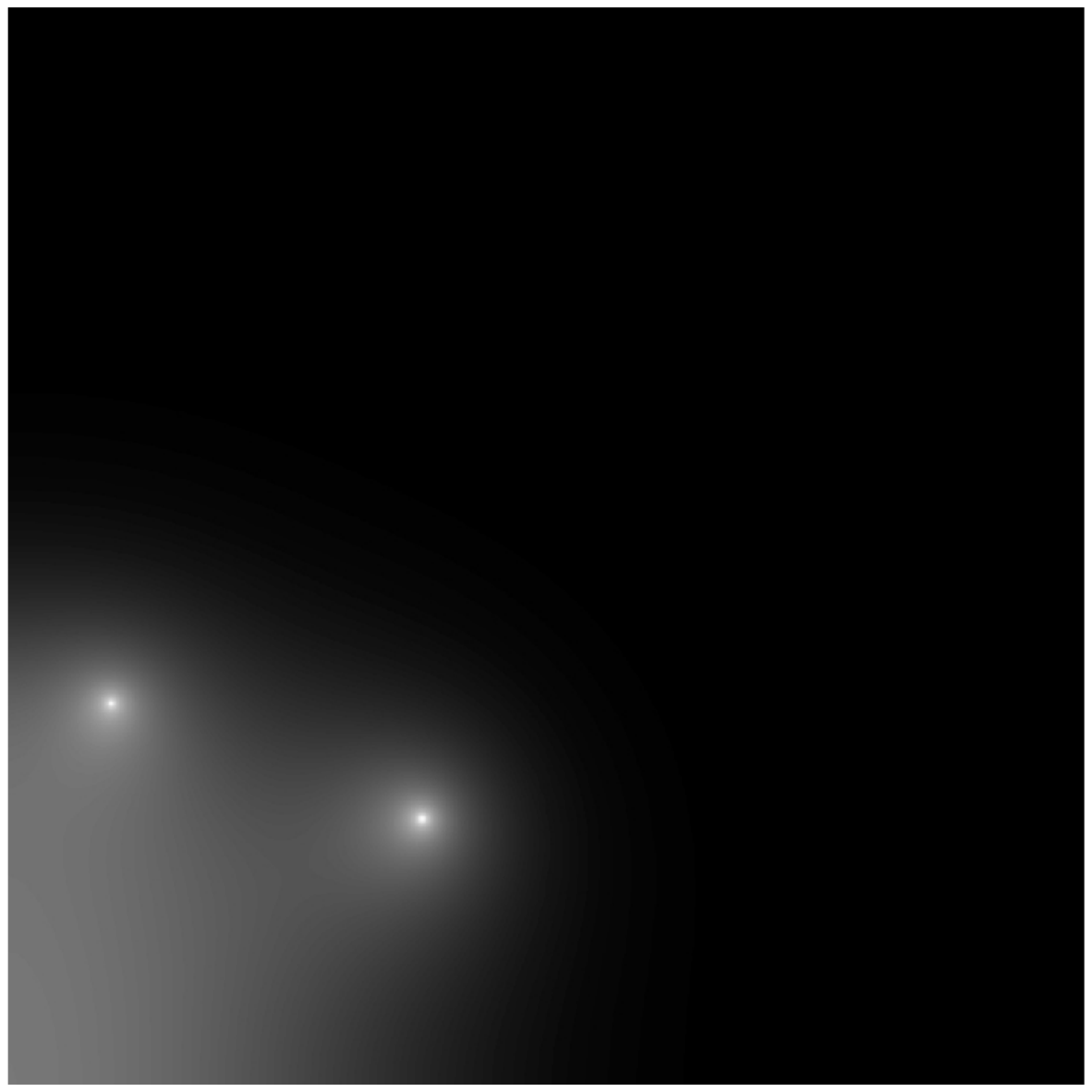}
         \caption{}
         \label{fig:3b}
     \end{subfigure}  
     \vspace{1pt}
	\caption{Distribution of the deficient reactant concentration, $C$,  at	$t=740$ (a) and $t=1000$ (b).  For conditions see the legend for Fig.\ref{kk2}. Lighter shading corresponds to lower concentration.} 
	\label{kk3}
\end{figure}

Figures \ref{kk2} and \ref{kk3} display the results of numerical simulations of the effect of heat losses on the breaking up of the flame in 2D.  The problem has also been explored by Martínez-Ruiz et al. [7] and Dejoan [8] through large-scale numerical simulations of a model accounting both for the gas thermal expansion, buoyancy and momentum loss.

\begin{figure}[h!]
	\centering
\captionsetup[subfigure]{labelformat=empty}
     \begin{subfigure}[b]{0.42\columnwidth}
         \centering
         \includegraphics[width=\columnwidth]{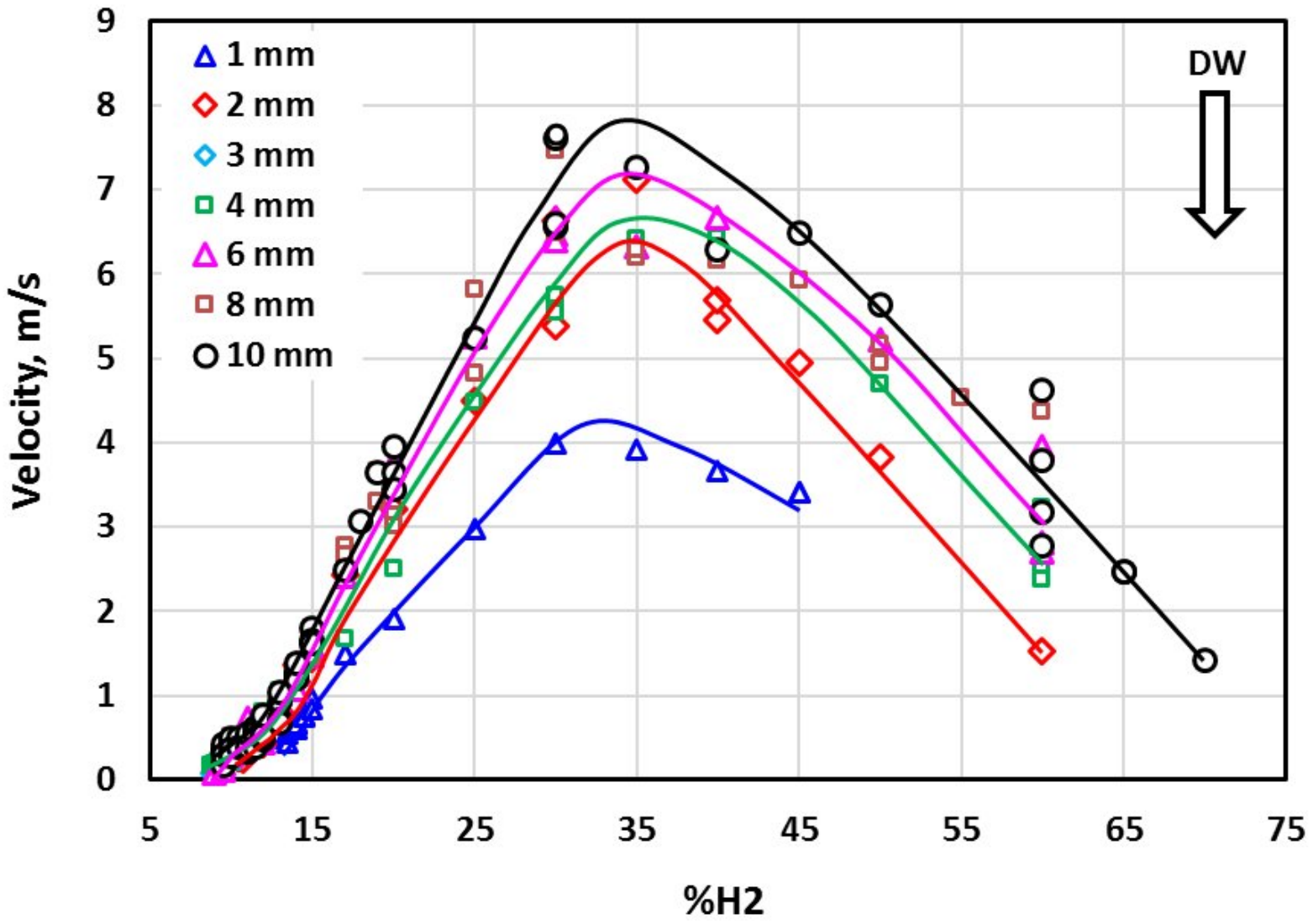}
         \caption{}
         \label{fig:5a}
     \end{subfigure}
     \begin{subfigure}[b]{0.42\columnwidth}
         \centering
         \includegraphics[width=\columnwidth]{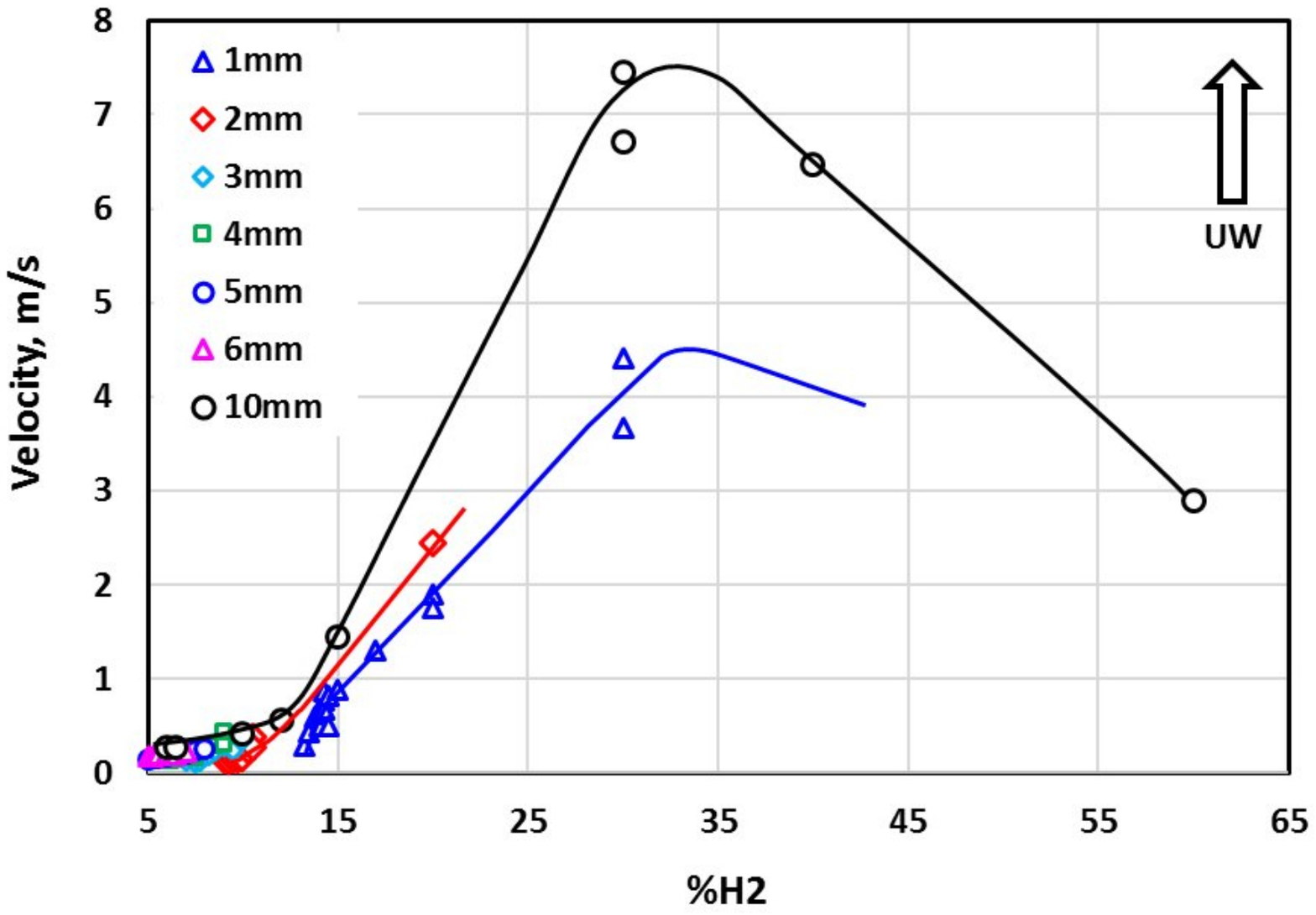}
         \caption{}
         \label{fig:5b}
     \end{subfigure} 
	\caption{Drift velocity for downward (DW) and upward (UW) propagation.  See also Kuznetsov et al. [9 Fig. 7]. The flame-ball regime corresponds to ultralean flames ($\% {H_2} < 20$). }
	\label{kk5}
\end{figure}

\begin{figure}[h!]
	\centering
	\includegraphics[width=0.6\columnwidth]{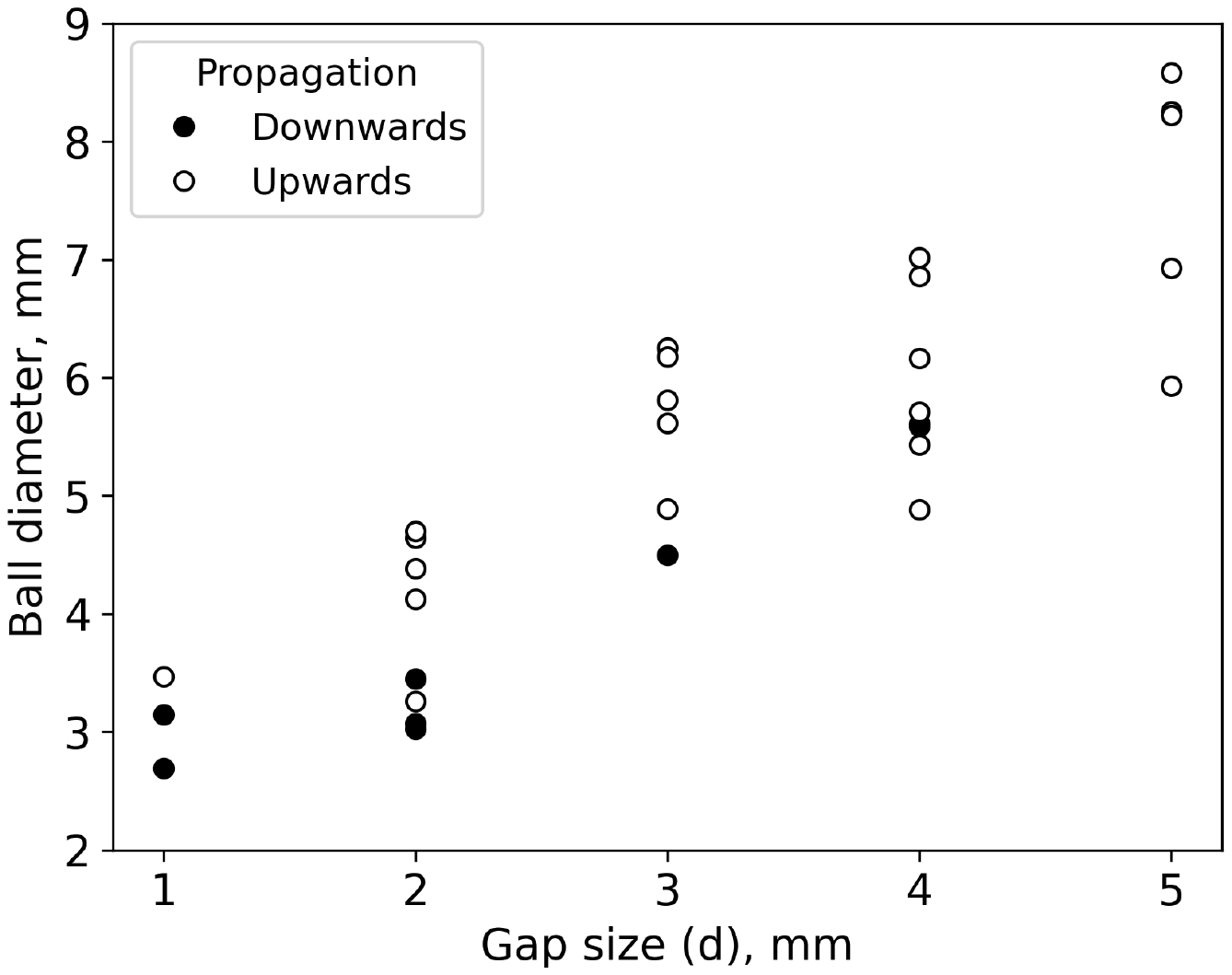}
\vspace{1pt}
 \caption{Flamelet diameter vs.\ gap size for one-headed flamelets.}
	\label{kk6}
\end{figure}

\begin{figure}[h!]
	\centering
	\includegraphics[width=0.5\columnwidth]{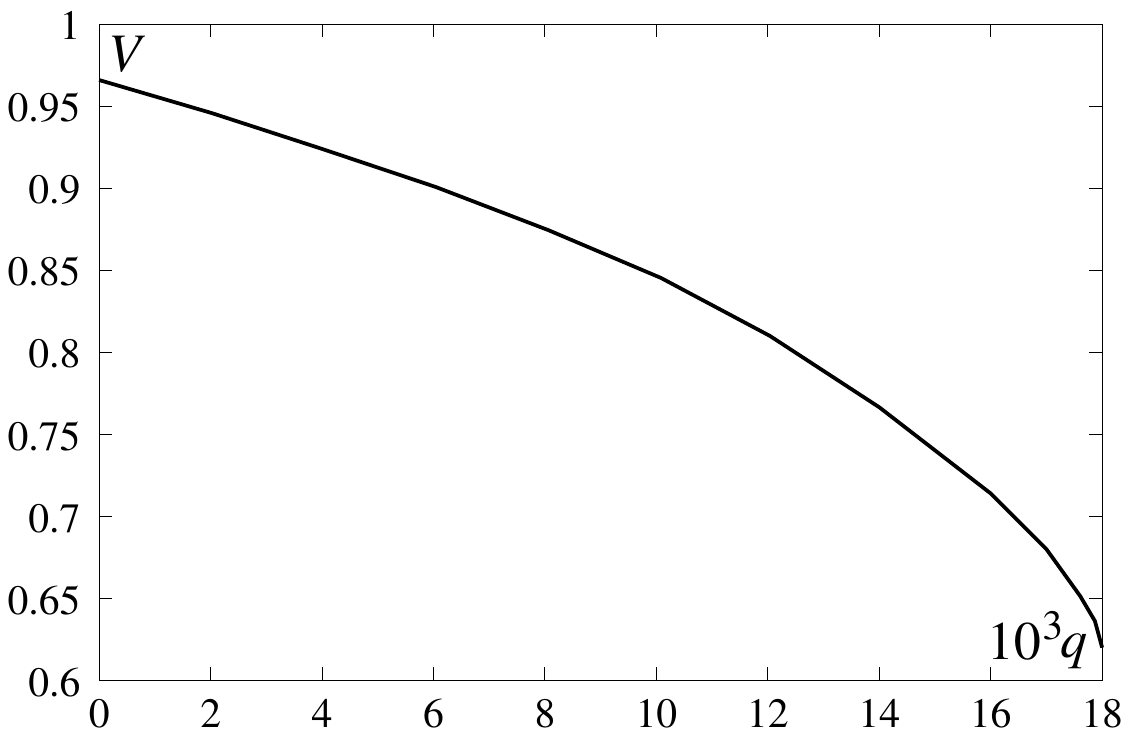}
 \vspace{-8pt}
	\caption{Drift velocity $V$ vs. heat loss intensity $q$ for a planar flame.}
	\label{kk4}
\end{figure}

Near the quenching point, $q=0.22$, higher heat losses lead to a smaller drift velocity and flamelet size (Fig. \ref{kk2}), which is in line with experimental findings (Figs. \ref{kk5}, \ref{kk6}). Note that $q=0.22$ significantly exceeds the quenching point $q=0.018$ for the planar flame, other conditions being identical (Fig.\ref{kk4}). 

\section{Reduction to a free-boundary problem}\addvspace{10pt}

   The fingering patterns of Figs.\ref{kk1}-\ref{kk3} are traces of self-drifting near circular reactive spots. They may be regarded as a 2D version of familiar self-drifting 3D flame-balls [10, 11] (see also comment ii) of Sec. 5).  This observation in turn suggests the possible existence of individual self-drifting 2D flame-balls described by time-independent reaction-diffusion equations,
\begin{align}
	&
	V\dfrac{\partial T}{\partial \eta}
	=
	\dfrac{\partial^2 T}{\partial \xi^2}
	+
	\dfrac{\partial^2 T}{\partial \eta^2}
	+
	(1-\sigma)\Omega
	-
	q(T-\sigma),\label{eq4}
	\\
		&
	V\dfrac{\partial C}{\partial \eta}
	=
	\frac{1}{Le}
	\left(
	\dfrac{\partial^2 C}{\partial \xi^2}
	+
	\dfrac{\partial^2 C}{\partial \eta^2}
	\right)
	+
	(1-\sigma)\Omega,\label{eq5}
\end{align}
subjected to boundary conditions,
\begin{equation}
	T(\xi^2+\eta^2\to\infty)=\sigma,
	\;
	C(\xi^2+\eta^2\to\infty)=1	.
\end{equation}
Here	$\xi=x$, $\eta=y+Vt$, and $V$ is an eigen-value of the problem.
For theoretical analysis it is technically advantageous to employ the familiar large-activation-energy $(N\gg 1)$ near-equidiffusive $(Le^{-1}-1\ll 1)$ approach where the reaction-diffusion system \eqref{eq4} and \eqref{eq5} is reduced to a free-boundary problem [12]. There, the reaction rate   $\Omega$ transforms into a localized source distributed along the interface	$\eta=R(\theta)\cos{\theta}$, $\xi=R(\theta)\sin{\theta}$. Eqs. \eqref{eq4} \eqref{eq5} translate into the familiar set of equations for the reduced temperature	$\Theta$  and excess-enthalpy $S$,
\begin{align}
	&
	V\dfrac{\partial \Theta^{(0)}}{\partial \eta}
	=
	\dfrac{\partial^2 \Theta^{(0)}}{\partial \xi^2}
	+
	\dfrac{\partial^2 \Theta^{(0)}}{\partial \eta^2},\label{eq7}
	\\
	&
	V\dfrac{\partial S}{\partial \eta}
	=
	\dfrac{\partial^2 \left(S-\alpha\Theta^{(0)}\right)}{\partial \xi^2}
	+
	\dfrac{\partial^2 \left(S-\alpha\Theta^{(0)}\right)}{\partial \eta^2}
	-
	\nu\Theta^{(0)}
\end{align}
Here	$\Theta=(T-\sigma)/(1-\sigma)$; $S=(\Theta+C-1)\beta$; the Zeldovich number $\beta=(1-\sigma)N$, is
 assumed to be large; $\alpha=\left(Le^{-1}-1\right)\beta$, $\nu=q\beta$		are assumed to be finite; $	\Theta^{(0)}=\Theta(\beta\to\infty)$.
 
At the reactive interface, $\xi^2+\eta^2=R^2(\theta)$, the following conditions are held,
\begin{align}
	&
	\left[S\right]^+_-=0, \;
	\Theta^{(0)}=1,\label{eq9}
	\\
	&
	\left[	\dfrac{\partial S}{\partial n}\right]^+_- 
	= 
	\alpha
	\left[	\dfrac{\partial \Theta^{(0)} }{\partial n}\right]^+_- ,
	\\
	&
	\left[	\dfrac{\partial \Theta^{(0)} }{\partial n}\right]^+_- 
	=-
	\left(1+\dfrac{1}{R^2}\left(\dfrac{\partial R}{\partial \theta}\right)^2\right)^{1/2}
	\exp{\left(\dfrac{S}{2}\right)}.
\end{align}
Here	$\partial /\partial n$  is the normal derivative.  
And,
\begin{align}
	&
	\Theta^{(0)} \to 0, \;
	S \to 0,
\end{align}
at $\xi^2+\eta^2\to\infty$. Within the flame-ball interface the deficient reactant is assumed to be fully consumed,
\begin{equation}
	C\left(\xi^2+\eta^2\le R^2(\theta)\right)=0.\label{eq13}
\end{equation}
Finally we observe that Eq. \eqref{eq7} allows for the exact solution which in polar coordinates ($\eta=r\cos{\theta},  \xi=r\sin{\theta}$) reads, [13, 14],
\begin{align}
    &
    \Theta^{(0)}(r,\theta)=\sum_m \exp{\left(kr\cos{\theta}\right)}\cos{\left(m\theta\right)}\cdot\nonumber\\
    &
    \cdot\left[A_m K_m(kr)+B_m I_m(kr)\right]+H.\label{eq:new14}
\end{align}
Here $K_m$, $I_m$ are modified Bessel functions of first and second kind, $k=V/2$, and $A_m$, $B_m$, $H$ are parameters to be determined by conditions at $r=0$, $r=\infty$ and $r=R(\theta)$.

\section{Reduction to a 1D model}
\subsection{General}
\addvspace{10pt}

\begin{figure}[h!]
	\centering
	\includegraphics[width=0.5\columnwidth]{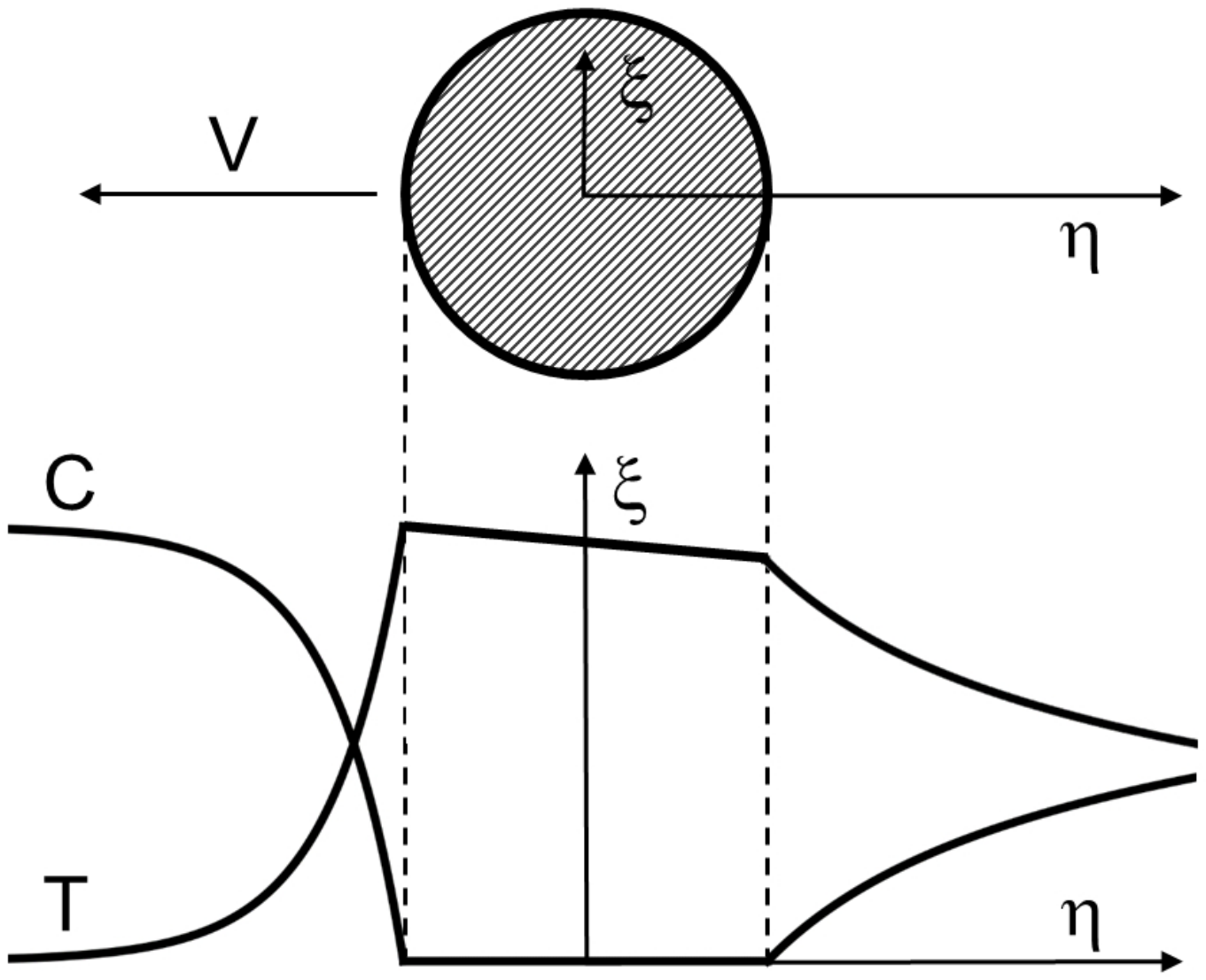}
    \vspace{0.3cm}
	\caption{Up: Top sketch view of the flamelet in a plane parallel both to the paper and to the plates of the Hele-Shaw cell. Down: Profiles of temperature, $T$, and concentration, $C$, through the center of the flame-ball.}
	\label{kk7}
\end{figure}

   The 2D free-boundary problem \eqref{eq7} - \eqref{eq13} is still too difficult for a straightforward analytical treatment.
   We therefore propose to consider its low-mode allocation-like reduction to a 1D model that we believe will keep enough contact with the original 2D formulation.
   More specifically, in the 1D approach $\Theta, C, S$ are projected on the  $\eta$ - axis running through the center of the flame ball (Fig. \ref{kk7}). 
   There the interior of the flame-ball covers the interval	$-R<\eta<R$. In the expansion \eqref{eq:new14} we will keep only the leading order terms corresponding to $m=0$. Moreover, we will set $\theta=0$ for $\eta>0$ and $\theta=\pi$ for $\eta<0$. Conditions \eqref{eq9} - \eqref{eq13} on the flame-ball interface	$\eta=\pm R$	and $\eta=\pm\infty$  thus become,
\begin{align}
	&
	\left[S\right]^+_-=0,
 \;
	\Theta^{(0)}=1,\label{eq14}
	\\
	&
	\left[\dfrac{\partial \Theta^{(0)}}{\partial \eta}\right]^+_-
	+
	\exp{\left(\dfrac{S}{2}\right)} =0,\label{eq15}
	\\
 	&
	\left[\dfrac{\partial S}{\partial \eta}\right]^+_- 
	=
	\alpha
	\left[\dfrac{\partial \Theta^{(0)}}{\partial \eta}\right]^+_-,\label{eq16}
 \end{align}
 \begin{align}
	&
	\Theta^{(0)}\left(\eta=\pm \infty\right) =0,
 \;
	S\left(\eta=\pm \infty\right)=0.
\end{align}
And
\begin{equation}\label{key}
	C\left(-R<\eta<R\right)=0.
\end{equation}
For the leading order approximation, $\beta\gg 1$,
\begin{align}
	&
	\Theta^{(0)}\left(\eta>R\right)
	=
	\exp{\left[k(\eta-R)\right]}
	\dfrac{K_0(k\eta)}{K_0(kR)},\label{eq19}
	\\
	&
	\Theta^{(0)}\left(\eta<-R\right)
	=
	\exp{\left[k(\eta+R)\right]}
	\dfrac{K_0(-k\eta)}{K_0(kR)},
\\
	&
	\Theta^{(0)}\left(-R<\eta<R\right)=1,
	\\
	&
	C^{(0)}\left(\eta\right)
	=
	1-\Theta^{(0)}\left(\eta\right).	
\end{align}
For the higher order uniform approximation needed for evaluation of $S=(\Theta+C-1)\beta$,
\begin{align}
	&
	\Theta\left(\eta>R\right)
	=
	\left[1+\beta^{-1}S(R)\right]\cdot
	\nonumber
	\\
	&
	\cdot\exp{\left[k(\eta-R)\right]}
	\dfrac{K_0\left( (k^2+q)^\frac{1}{2}\eta\right)}
	{K_0\left( (k^2+q)^\frac{1}{2} R\right)},
	\\
	&
	\Theta
 \left(\eta<-R\right)
	=
	\left[1+\beta^{-1}S(-R)\right]\cdot
	\nonumber\\
	&
	\cdot\exp{\left[k(\eta+R)\right]}
	\dfrac{K_0\left(-(k^2+q)^\frac{1}{2}\eta\right)}{K_0\left((k^2+q)^\frac{1}{2}R\right)},
	\\
	&
	\Theta\left(-R<\eta<R\right)
	=
	1+\beta^{-1}S\left(-R<\eta<R\right),\\
	&
	S\left(-R<\eta<R\right)=A+B\exp{(k\eta)}I_0(k|\eta|)\nonumber
        \\
        &
        \mspace{145mu}
        -\nu\eta/2k,
\end{align}
where,
\begin{align}
	&
	A
	=
	\left[S(R)+S(-R)\right]
	\nonumber
	\\
	&
	-
	\dfrac{1}{2}
	B
	I_0(kR)
	\left[
	\exp{(kR)}
	+
	\exp{(-kR)}
	\right]
	\\
	&
	B=
	\dfrac{\nu R+S(R)-S(-R)}
	{k I_0(kR)\left[\exp{(kR)}-\exp{(-kR)}\right]}
	,
	\\
	&
	C(\eta>R)
	=
	1
	-
	\exp{\left[k Le (\eta-R)\right]}
	\dfrac{K_0(k Le \eta)}{K_0(k Le R)},
	\\
	&
	C(\eta <\! -R)
	=
	1
	-
	\exp{\left[k Le (\eta+R)\right]}
	\dfrac{K_0(-k Le \eta)}{K_0(k Le R)},
	\\
	&
	C(-R<\eta<R)=0.	\label{eq31}
\end{align}

The uniform approximation employed in the above equations allows meeting the boundary conditions $\Theta(\eta\to\infty)=0$, $S(\eta\to\infty)=0$, behind the self-drifting flamelet, where $\eta \sim \beta$.

\vspace{10pt}

\subsection{Results}

\addvspace{10pt}

\begin{figure}[h!]
  \centering
  \begin{minipage}{.45\textwidth}
    \includegraphics[width=\columnwidth]{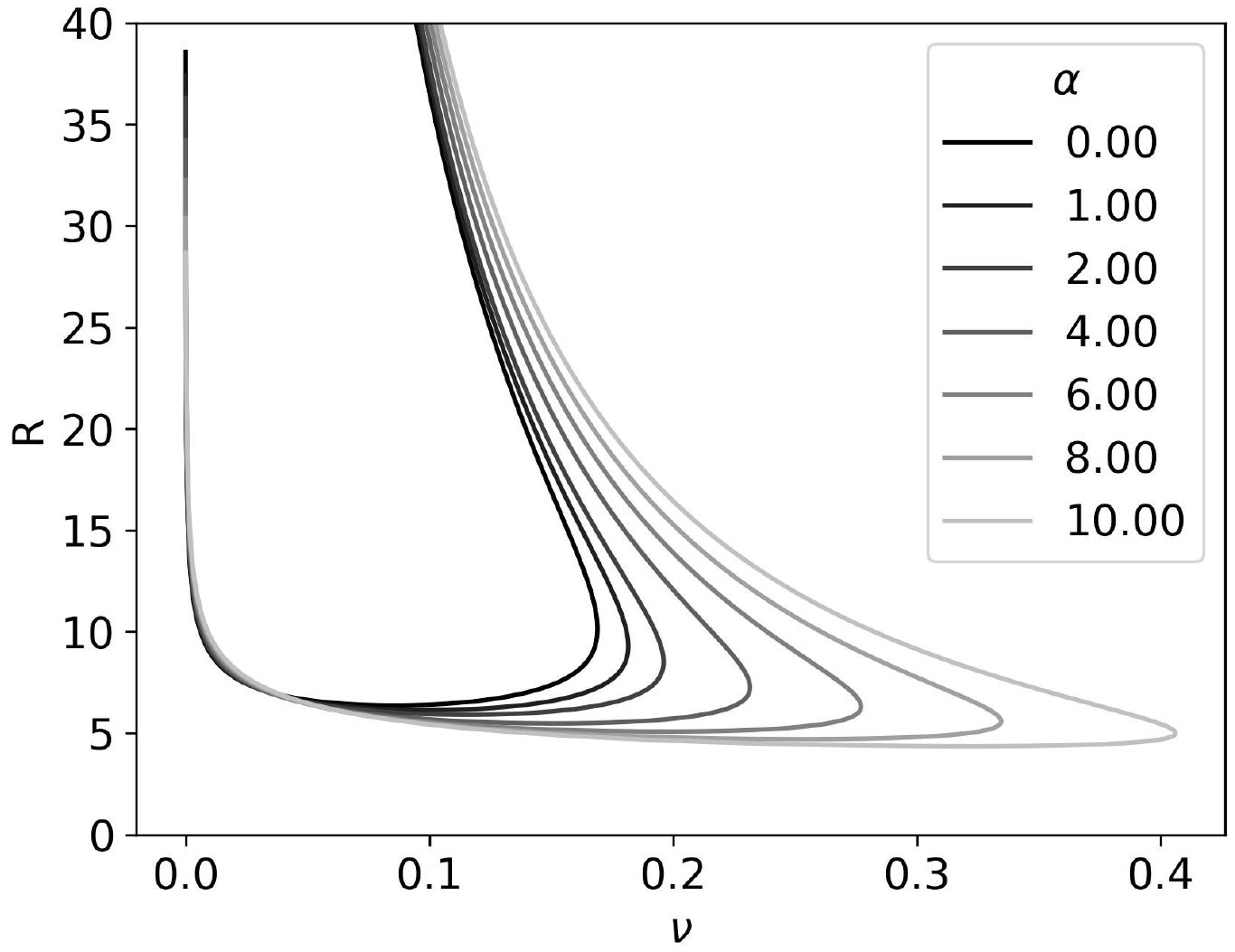}
 \vspace{-5pt}
 \caption{Locus of the flamelet radii, $R$, as a function of the heat losses, $\nu$, for different Lewis number parameters, $\alpha$.}
 \label{kk9}
  \end{minipage}\hspace{0.5cm}%
 \begin{minipage}{.45\textwidth} 
  \centering
  \includegraphics[width=\columnwidth]{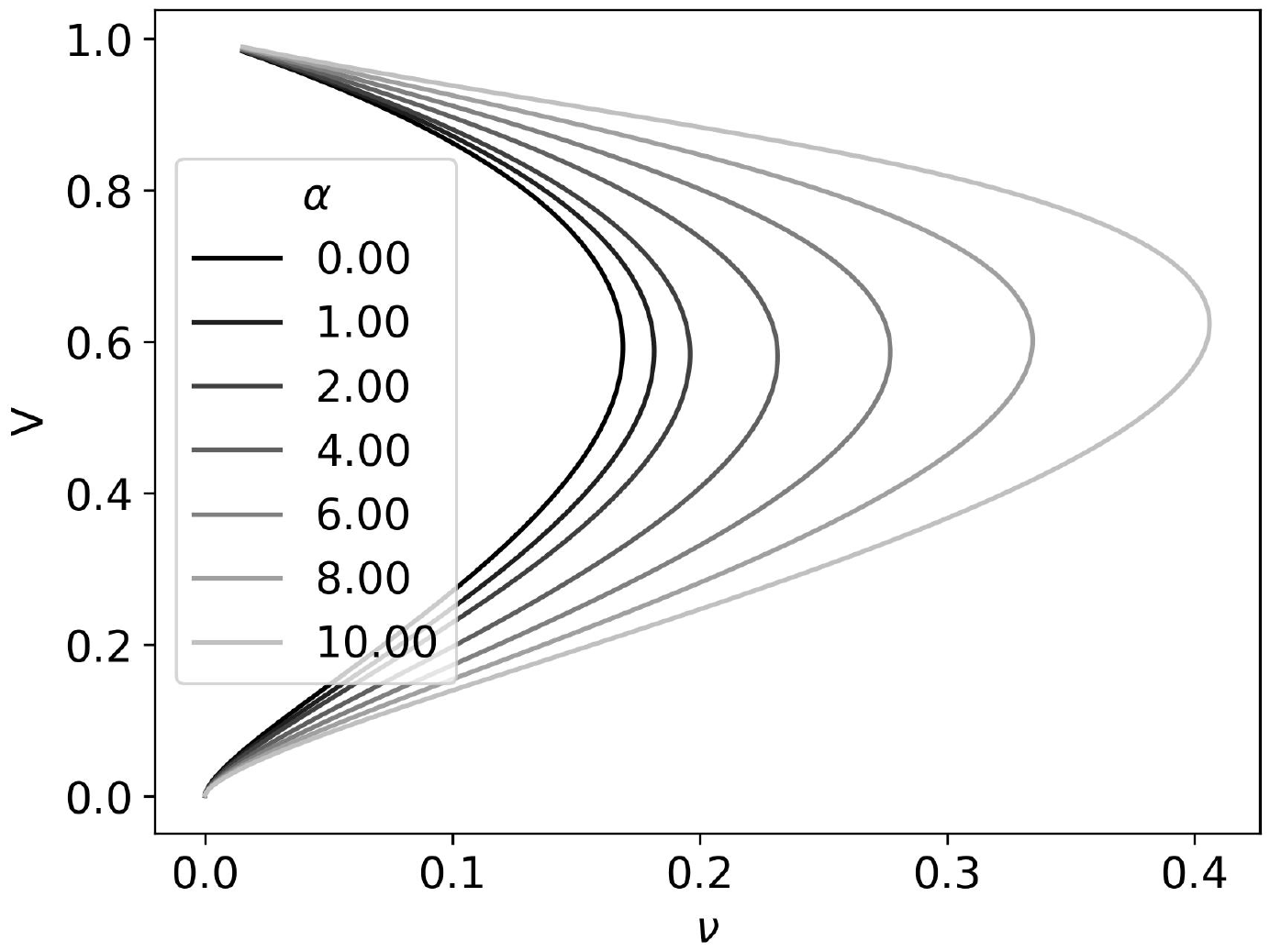}
  \vspace{-5pt}
  \caption{Locus of the velocities, $V$, as a function of the heat losses, $\nu$, for different Lewis number parameters, $\alpha$.}
  \label{kk8}
  \end{minipage}
\end{figure}

\begin{figure}[h!]
	\centering
	\includegraphics[width=0.5\columnwidth]{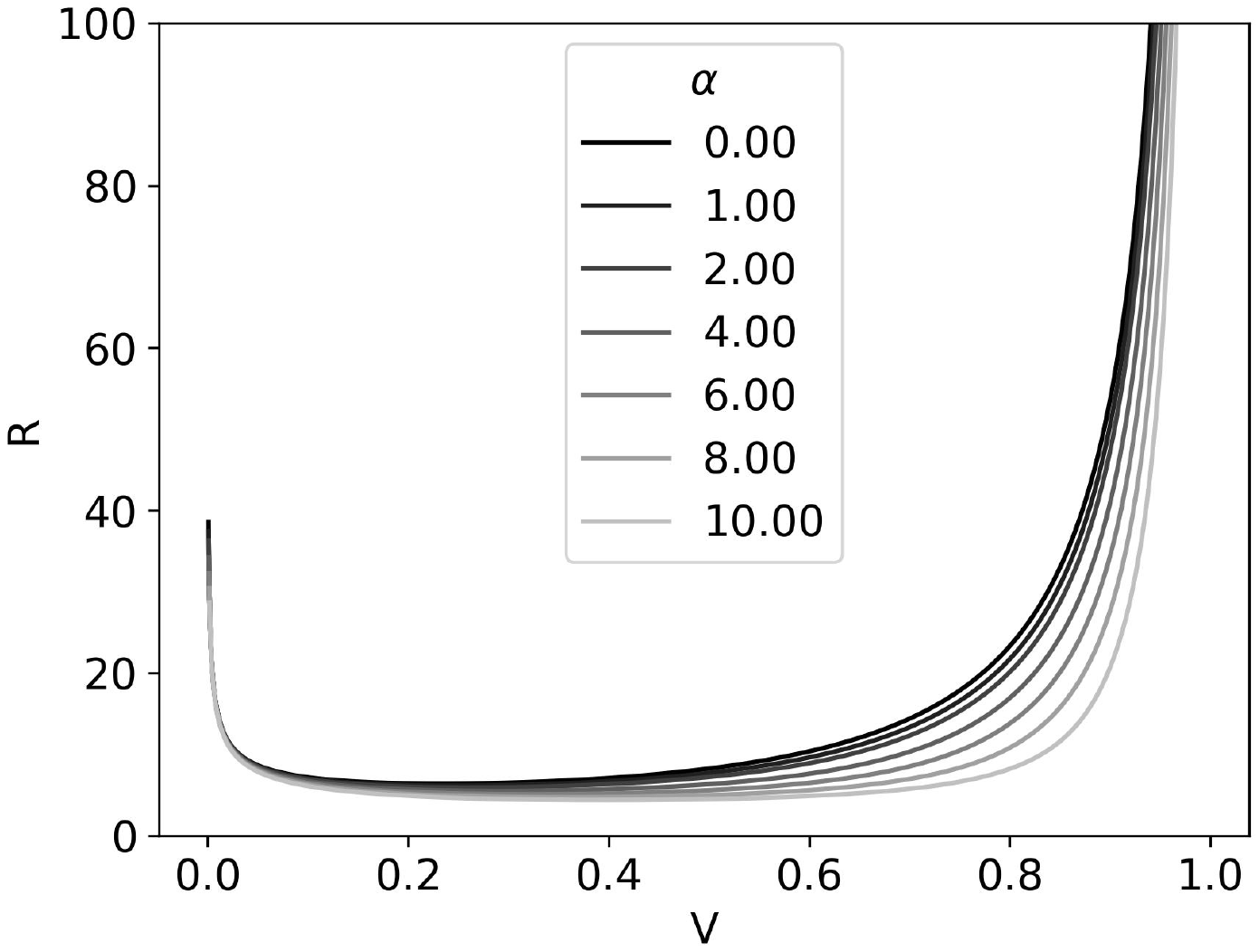}
  \vspace{-5pt}
	\caption{Locus of the flamelet radii, $R$, as a function of velocity, $V$, for different Lewis number parameters, $\alpha$.}
	\label{kknew9}
\end{figure}

Eqs. \eqref{eq19} - \eqref{eq31} meet the continuity conditions \eqref{eq14} at	$\eta=\pm R$.
Substituting Eqs. \eqref{eq19} - \eqref{eq31} into jump conditions \eqref{eq15}, \eqref{eq16} one ends up with four algebraic relations for four unknown parameters	$V=2k$, $R$,	$S(R)$, $S(-R)$ as functions of	$\alpha$, $\nu$, $\beta$ (Figs.  \ref{kk9}, \ref{kk8}, \ref{kknew9}). More technical details and the numerical procedure employed for the emerging algebraic problem are presented in the earlier version of the paper [15].

One can draw several conclusions from the results obtained.

\begin{enumerate}[\hspace{-8pt} (a)]

 \item Parameter $\alpha$ strongly affects $V(\nu)$, $R(\nu)$ - dependencies.
  The heat loss intensity at the quenching (turning) point, $\nu_q$,
  increases with the decrease of the Lewis number, $Le \simeq 1-\alpha/\beta$.

 \item  At $\nu<\nu_q$ the multiplicity of  $V(\nu)$, $R(\nu)$ is observed. 
   A similar non-uniqueness is known to occur in planar ($R=\infty$)
   non-adiabatic flames which leads to the $\alpha$-independent relation
   between  $V$ and $\nu$, $V^2\ln{\left(1/V\right)} =\nu$, [16].

 \item  Once the solutions for	$V(\nu)$, $R(\nu)$ for a certain $\alpha$ are
   available, one can plot spatial profiles of all state variables involved.
   For brevity we depict here only $\Theta(\eta)$ for the upper and lower
   branches of	$V(\eta)$ (Figs. \ref{kk10}, \ref{fig:teh1lga}).
   A significant drop of temperature	$\Theta$ at the rear side ($\eta=R$) of the self-drifting flame-ball might explain the horseshoe shape of the advancing flamelet.  The latter is often perceived as a local extinction (opening ) of the front.	
\begin{figure}[h!]
  \centering
    \begin{minipage}{.45\textwidth}
      \includegraphics[width=\columnwidth]{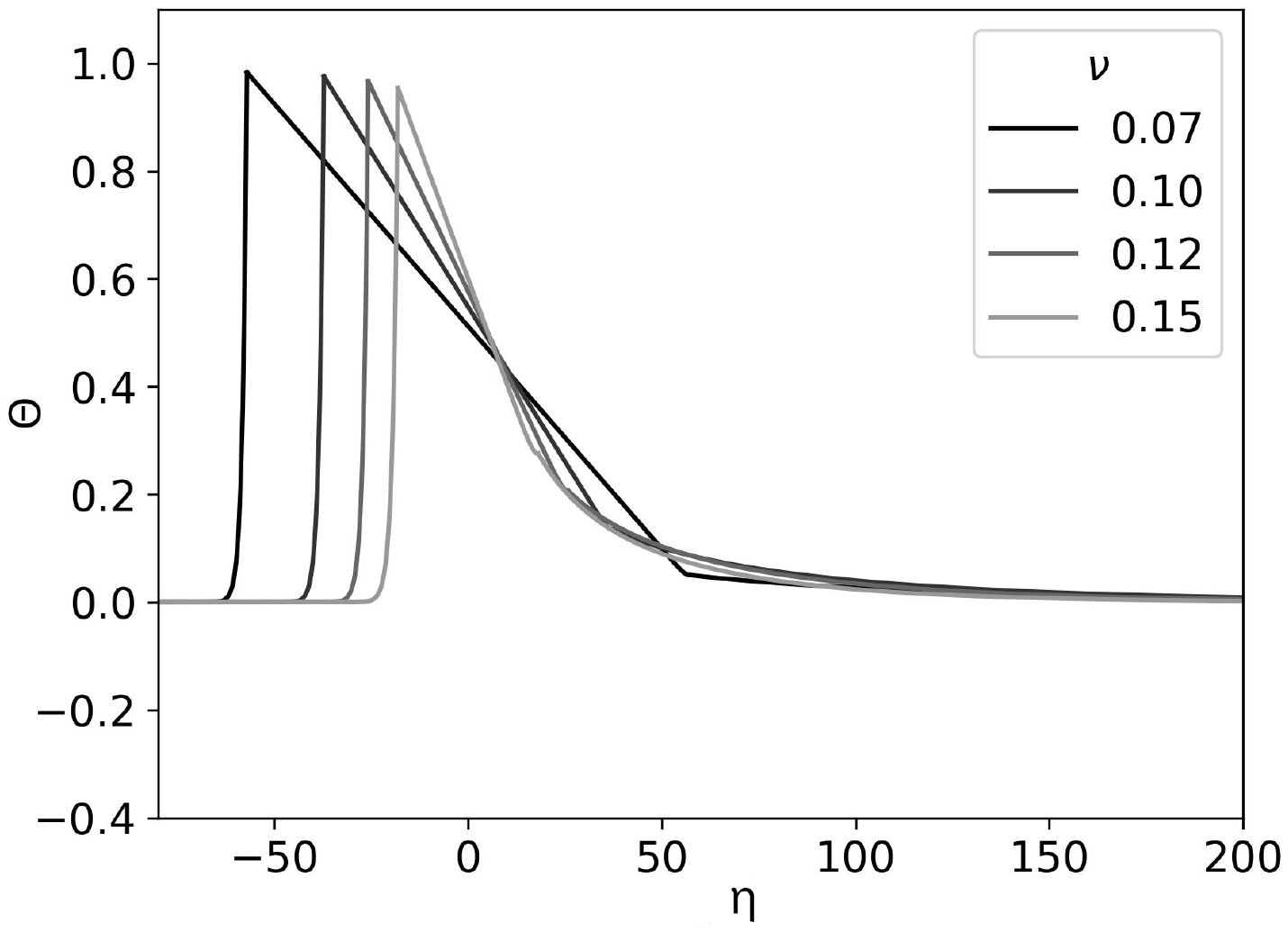}
      \vspace{-5pt}
      \caption{Profiles of $\Theta$ for $\alpha=1$, $\beta=10$ and different values of $\nu$ for the upper branch of $V(\nu)$. }
      \label{kk10}
    \end{minipage}\hspace{0.5cm}%
   \begin{minipage}{.45\textwidth} 
     \centering\
     \includegraphics[width=\columnwidth]{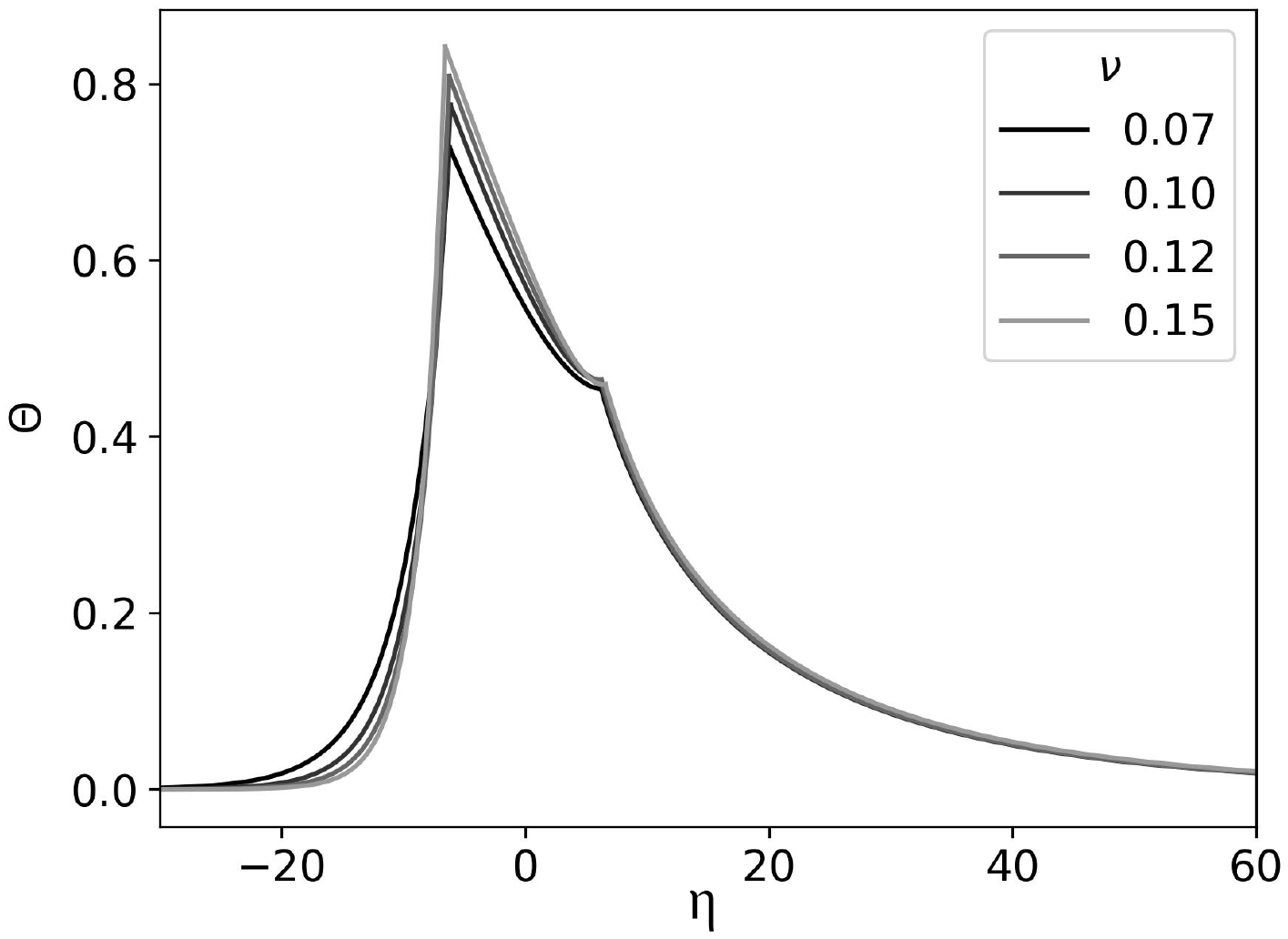}
     \vspace{-5pt}
     \caption{Profiles of $\Theta$ for $\alpha=1$, $\beta=10$ and different values of $\nu$ for the lower branch of $V(\nu)$. }
     \label{fig:teh1lga}
   \end{minipage} 
\end{figure}
 
\end{enumerate}

\section{Concluding remarks}\addvspace{10pt}

\begin{enumerate}[\hspace{-8pt} (i)]

\item  In constructing a workable approximate solution we truncated the series involving an infinite number of terms, as in Eq. \eqref{eq:new14}, setting $m$      at zero, to deal with sums involving only a finite number of terms.

\item  

  There is an important distinction between the 3D and effectively 2D
  situation typical of large aspect ratio Hele-Shaw cells.
  The 3D case allows for stationary spherical flame-balls which may bifurcate into a self-drifting mode [10, 11].
  In the 2D case, the non-drifting circular flame-balls are ruled out. They cannot meet boundary conditions at infinity (due to the logarithmic tail of the associated concentration profiles [7]).
  So, in narrow gaps 2D self-drifting flame-balls should emerge not as a bifurcation but rather as the only way for the 2D flame-balls to exist.

\item In the present study the pattern of the flame disintegration is
  controlled by the heat loss intensity $q=(\pi\ell_{th}/d)^2$, which in
  turn may be controlled by the width of the Hele-Shaw gap ($d$) or
  the mixture composition (\% $H_2$ or $\ell_{th}$) [4, 9, 17].

\item  The reaction-diffusion model of Sec. 2 does not seem capable of reproducing two-headed flamelets. The latter is often observed experimentally [4, 18]
  as well as in simulations of more sophisticated models that consider the burned gas thermal expansion [7, 19]
  and are therefore susceptible to Darrieus-Landau instability.  At the moment the issue of two-headed flamelets  remains unexplained.

\item  The present work is devoted to estimates of the propagation velocity of an individual flame-ball in a flat channel.  It would be interesting to extend the analysis over the collective propagation of a group of flame-balls appearing in Figs. \ref{kk1}-\ref{kk3}, and where the flamelets are competing for common fuel and mutual heating of one another.  A mean-field type of approach as the developed by D'Angelo and Joulin [20] or Williams and Grcar [21] seems particularly promising.

\item In the present study flame-balls emerge in gaseous premixtures as an extreme case of diffusive-thermal instability where invariably	$Le<1$.  A similar pattern is observed in smoldering burning of thin solid sheets with and without imposed air flows [22-28].
  There, similar to the gaseous systems, the effective Lewis number is considerably below unity [28].  It would be interesting to extend the 1D approach of this paper also to the smoldering problem.
\end{enumerate}

\acknowledgement{Declaration of competing interest}\addvspace{10pt}

The authors declare that they have no known competing financial interests or personal relationships that could have appeared to influence the work reported in this paper.

\acknowledgement{Acknowledgements} \addvspace{10pt}

   This research was supported in part by the US-Israel Binational Science Foundation (Grant 2020-005).


 \footnotesize
 \baselineskip 9pt


\end{document}